\documentclass[prl,superscriptaddress,twocolumn,amsmath,aps]{revtex4}
\usepackage{graphicx} 
\usepackage[utf8]{inputenc}
\usepackage{color}

\begin{document} 
\title{Characterizing order in amorphous systems}

\author{François Sausset}
\email{francois.sausset@lptms.u-psud.fr}
\affiliation{Department of Physics, Technion, Haifa 32000, Israel}
\affiliation{Univ. Paris-Sud \& CNRS, LPTMS, UMR8626,  B\^{a}t.~100, 91405 Orsay, France.}

\author{Dov Levine}
\affiliation{Department of Physics, Technion, Haifa 32000, Israel}

\date{\today}
\begin{abstract}
	We measure and compare three correlation lengths proposed to describe the extent of structural order in amorphous systems. In particular, the recently proposed ``patch correlation length'' is measured as a function of temperature and fragility and shown to be comparable with other measures.  In addition, we demonstrate that the patch method also allows us to characterize the symmetries of the local order without any \emph{a priori} knowledge of it.
\end{abstract}

\maketitle

Amorphous solids are ubiquitous, and include glasses, sand piles, emulsions, and colloidal systems~\cite{Debenedetti:2001, Jaeger:1996, Berthier:2010p919}.  Such systems appear at best to possess short-range order~\cite{Berthier:2010p919, Biroli:2008}, in stark contrast to crystals and quasicrystals, whose order may well extend over an entire macroscopic sample.
One striking common feature of amorphous systems is slow dynamics: thermal systems become increasingly sluggish upon cooling~\cite{Debenedetti:2001, Berthier:2010p919}, showing a remarkable rapid rise in viscosity, while athermal systems jam when compressed~\cite{Jaeger:1996, Liu:1998p430, Torquato:2010}. While it is easy to understand how motion in an ordered system is difficult, it is harder to rationalize such slow dynamics in amorphous systems, which leads to a fundamental question: Is the extraordinary slowing-down related to some subtle growing structural order?  This question is the subject of much current activity~\cite{Bouchaud:2004, Montanari:2006, Biroli:2008, Kurchan:2009p642, Kurchan:2010p793, Berthier:2010p919, Sausset:2010, Mosayebi:2010}, and its resolution would lead to a better understanding of glassy phenomena, and could facilitate the testing of various theories of the glass transition~\cite{Berthier:2010p919}.

Classical methods used to probe structural order show no dramatic change in the structure of a material as it approaches the glass transition~\cite{Berthier:2010p919}. These methods are based on scattering experiments and only give access to two-point correlations, such as the pair correlation function and the structure factor. Thus, if there is any (``hidden'') order present in glassy systems, it would have to be very subtle, and inaccessible to conventional techniques.

A major obstacle in measuring the extent of order in conventionally amorphous systems is our lack of knowledge of what that order might be. This is unlike more familiar systems in condensed matter where we identify the order parameter at the outset. Can we, in the absence of a well-defined order parameter, define a characteristic correlation length?
Since the nature of a possible order in a glassy phase is not known, a good definition of a correlation length should assume nothing about the nature of the order it seeks to quantify.
That is, structural correlation lengths should be defined in an \emph{order agnostic} way.  
Such a procedure would equally well aid in the identification of possible hidden order in other, non glassy systems, such as colloidal suspensions.

Interest in the glass transition has spurred several recent proposals of structural correlation lengths in supercooled liquids~\cite{Biroli:2008, Sausset:2010, Mosayebi:2010}. 
For example, the ``point-to-set'' length~\cite{Bouchaud:2004, Montanari:2006, Biroli:2008} assumes nothing about the nature of the developing order, but has the disadvantage of being difficult to implement experimentally since it requires freezing a portion of the system in its equilibrium configuration while letting the rest evolve.  The proposal of Ref. \cite{Sausset:2010} relies on {\it a-priori} knowledge of the developing order, which is the case for the frustrated system studied here.  Although it is not order-agnostic, we use it in this Letter as a benchmark for comparison.

Recently, a novel, generic method to define and measure a static correlation length~\cite{Kurchan:2009p642, Kurchan:2010p793} has been proposed, based on the frequency of occurrence of motifs (``patches'') of various sizes.  
The main idea is to compute the entropy of patches of a given size $R$ appearing in a system~\footnote{This is given by $S_R = -\sum P_{j} \log P_{j}$, where $P_{j}$ is the frequency of occurrence of the $j^{th}$ patch of size $R$.}, and then to look at how this entropy scales with $R$.  This scaling yields information about the extent of order in the whole system, even if no order parameter has been identified, and allows us to define a  ``patch correlation length''.

There are two main points to this paper: First, we calculate the patch length for the frustrated atomic system of Ref.~\cite{Sausset:2010} as a function of temperature, and compare it to other proposed correlation measures.  Second, we demonstrate how patch entropy considerations may be used to determine the nature of the developing order. 
 
Although the patch entropy method does not require patches of any specific shape, we shall find it convenient in what follows to consider spherical patches of radius $R$.
To get a feeling for the behavior of the patch entropy, we first consider a perfect undefected crystal.  It is easy to check that for $R$ greater than some radius (roughly the size of the unit cell), the number of possible patches is constant, independent of $R$, and therefore so is the patch entropy $S(R)$\cite{Kurchan:2009p642,Kurchan:2010p793}. The opposite extreme is a totally random system, where the number of patches is exponential in the volume $V(R)$ of the patch, leading to $S(R) \propto V(R)$ - that is, it is extensive in patch size. For a quasicrystal, $S(R) \propto \log R$, i.e., the entropy is subextensive in patch size, albeit greater than for a crystal.

Next, consider a system ordered only up to a finite distance $\xi$. It might be expected that in such a case, the system would, on short scales, organize into low energy patches which are present in the ground state - this is what happens, for example, in a polycrystal.  In this case, small patches ($R<\xi$) will occur often, and the patch entropy will be subextensive in $R$ for $R<\xi$.  On the other hand, since any given large patch will occur exponentially infrequently, $S(R) \sim V(R)$ for $R>\xi$. (For the polycrystalline case, for example, patch congruence would require the identical configuration of grain boundaries.) The scale $\xi$ where the behavior of $S(R)$ changes behavior from subextensive to extensive was proposed~\cite{Kurchan:2009p642, Kurchan:2010p793} as a measure of the correlation length of the system; this is the patch correlation length $\xi_S$.  In practice, it is easier to measure this length by measuring the inverse slope of the extensive regime of the entropy; this is what we do in the following.  In particular, one defines a correlation volume $V^* \sim \xi_S^d$ ($d$ being the space dimension) by noting that in the extensive regime, $S(R) \propto \frac{V(R)}{V^*}$. (Since this is only a proportionality relation, we cannot fix the absolute value of $\xi_S$.)

In~\cite{Kurchan:2009p642, Kurchan:2010p793}, it was demonstrated that this proposal for defining order applies not only to systems whose ground states are crystalline or quasicrystalline, but also to other kinds of exotic ground states whose order is not detected by scattering. Thus, this method may reveal ``hidden'' order present in a system.  As argued in Ref. \cite{Kurchan:2010p793}, for a non-stationary system evolving in time, the relevant patches to be studied are derived from the density profile which persists after high frequency vibrations are averaged away.
This consideration is based on a separation of time scales, which is seen in the relaxation of glassy systems: rapid motions occur on the so-called $\beta$ relaxation time $\tau_\beta$, whereas the long time $\alpha$-relaxation ($\tau_\alpha$) corresponds to collective structural relaxation.  The rapid motions then average out over times $\tau$ such that $\tau_\beta \ll \tau \ll \tau_\alpha$, and we are left with a density profile on which we may measure the patch entropy.  

In what follows, we will analyze {\it static} configurations of the model described in~\cite{Sausset:2010} that are obtained by quenches to low temperatures. The use of static configurations is motivated by experimental systems for which snapshots are accessible while time averaging is difficult. In this case, however, no two patches are exactly identical, and we need to define a procedure for quantifying the ``distance'' between two patches.  For the purpose of computing the patch entropy, we shall consider as congruent two patches whose distance is less than some cutoff.  Our rationale is that two very similar patches from a static configuration would, could we observe them in time, evolve to the same average density profile.  In consequence, a single (sufficiently large) snapshot of a system should serve in lieu of time averaging. We suggest that choosing the cutoff from the value of the plateau seen between the $\beta$ and $\alpha$ regimes in the mean square displacement in glassy systems~\cite{Debenedetti:2001} should be analogous to averaging over a time corresponding to the end of this plateau.

The detailed procedure is as follows: (i) Select a particle $i$ in the system and consider the surrounding configuration contained in a sphere of radius $R$ centered on $i$ - denote this patch $\Pi_{i}(R)$; (ii) Superimpose this patch on $\Pi_{j}(R)$, the corresponding patch of another particle $j$, so that the centers coincide; (iii) For each particle in $\Pi_{i}(R)$, compute the distance $d$ from the nearest particle of the patch $\Pi_{j}(R)$.  If $d$ is less than a given threshold $d^*$, consider these two particles to be matches; (iv) Rotate the patch by some small angle $\delta \alpha$ and repeat step iii.  If, for some angle $\alpha$, each particle inside the (rotated) patch $\Pi_{i}(R)$ matches one in $\Pi_{j}(R)$, consider the two patches to be congruent~\footnote{This relation is not necessarily commutative, as in the case of two patches identical but for the addition of a single particle.}; (v) Build a congruence matrix $C$ whose element $c_{ij}$ equals $1$ if $\Pi_{i}$ and $\Pi_{j}$ are congruent, otherwise set $c_{ij} = 0$.  Next, symmetrize $C$: if $c_{ij} \neq c_{ji}$, set $c_{ij} = c_{ij} = 0$ (so that if $\Pi_{i}$ is congruent to $\Pi_{j}$ then the reverse is true as well); (vi) Construct equivalence classes by grouping patches such that all the patches in a class are congruent with at least one other member of the class~\footnote{The underlying equivalence relation is not known and is different from the congruence relation just described.}.  Intuitively, patches belonging to the same equivalence class look similar, while those belonging to different classes look different.

Having constructed the equivalence classes, we compute the associated patch entropy $S(R)$: 
\begin{equation}
	S(R) = - \sum_\alpha P_{\alpha}(R) \ln P_{\alpha}(R) 
\end{equation}
where the sum runs over equivalence classes, and $P_{\alpha}(R)$ is the probability of a patch to be in the $\alpha^{th}$ class.  We next examine the functional dependence of $S$ on $R$.
As discussed earlier, one can extract a correlation length either by identifying the crossover length where $S(R)$ goes from subextensive to extensive in $R$, or by measuring the slope of the extensive regime.

\begin{figure}
	\centering
		\includegraphics[width=8 cm,trim= 1cm 0cm 0.5cm 0cm]{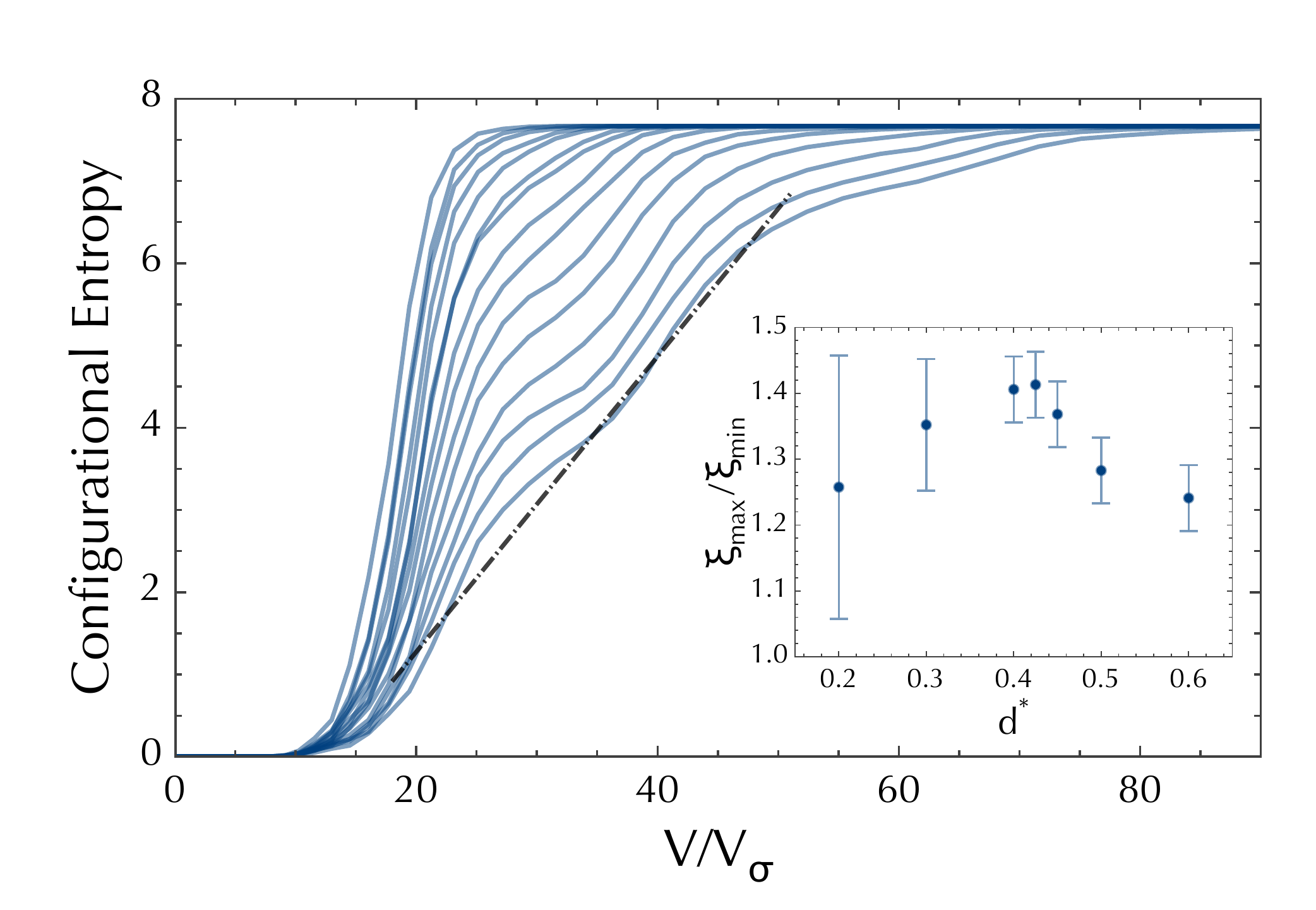}
\caption{Patch entropy as a function of patch volume: temperatures range from $T/T^* = 4.35$ (left) to $T/T^* = 0.59$ (right) for a system of $2138$ particles. $V_\sigma$ is the volume of a single particle and $T^*$ the temperature of the avoided crystallization (see Ref.~\cite{Sausset:2008b}). Note the three regimes of the entropy: constant and near zero for small sizes, then increasing linearly, and finally saturation.
The patch correlation length is extracted from the inverse of the slope of the linear (middle) regime, measured by a linear fit indicated by a dot-dash line for the lowest temperature.  Insert: ratio of extracted length scales for $T/T^* = 0.35$ and $T/T^* = 3.93$ for the less fragile system versus the distance threshold involved in patch comparison. In this article, the optimum value $d^*=0.425$ given by the maximum has been used.}
\label{fig:entropy}
\end{figure}

We will now apply these ideas to the model glassy system of Refs. \cite{Sausset:2008b,Sausset:2010,Sausset:2008d}, which consists of monodisperse Lennard-Jones particles placed on a two-dimensional negatively curved surface.  The system tends to have a local hexagonal order, whose spatial extension is frustrated due to the curvature of the surface.  Nevertheless, hexagonal or hexatic order persists over a temperature-dependent length scale, and may be quantified with an appropriate correlation function.  In this model, the dynamic fragility of the system can be tuned by varying the curvature of the surface~\cite{Sausset:2008b}.
\begin{figure}
	\centering
		\includegraphics[width=8 cm,trim= 1cm 0cm 0.5cm 0cm]{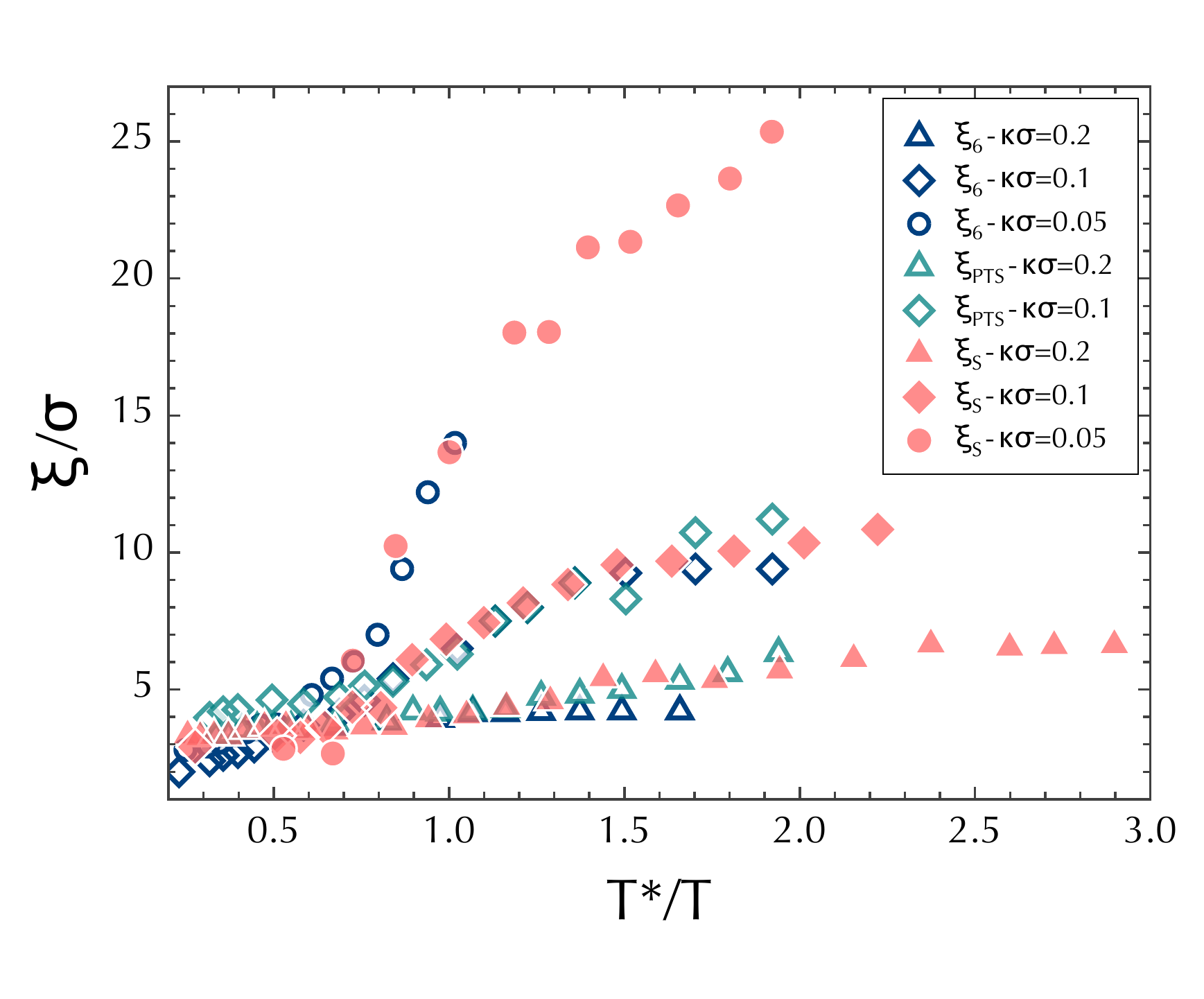}
\caption{The structural correlation lengths {\it vs.} temperature for different fragilities: $\xi_6$ (dark blue open symbols) denotes the hexatic length, $\xi_{\mathrm{PTS}}$ (light green open symbols) is the \emph{point-to-set} length, and $\xi_S$ (solid red symbols) is the patch correlation length. Errors bars are roughly of the same size as the symbols. The vertical axis corresponds to absolute values of $\xi_6$. $\xi_{\mathrm{PTS}}$ and $\xi_{S}$ have been linearly scaled but their raw values are of the same order of magnitude as $\xi_6$: $\xi_{\mathrm{PTS}}$ is multiplied by a factor between $0.56$ and $1.1$ (depending on the curvature), $\xi_S$ by a factor between $0.52$ and $0.65$, and both have been vertically shifted. The three different curvatures $\kappa$ correspond to three different fragilities of the system, with the smallest curvature being associated with the largest fragility. }
\label{fig:lengths}
\end{figure}

To evaluate the patch entropy of this system, we must first determine the relevant value of the threshold $d^*$.  As argued above, a reasonable choice for $d^*$ is the value of the mean particle displacement just after the plateau, a value that does not depend much on temperature.  One possible caveat is that this choice could yield poor ``contrast'', i.e., only little differentiation between systems at different temperatures (we expect an increase in the patch correlation length as temperature decreases). 

To check the ``contrast'', consider two typical snapshots of the system, one at low temperature and another at higher temperature, and evaluate the ratio of patch correlation lengths $\gamma(d^{*}) \equiv \xi_{\mathrm{low} T}/\xi_{{\mathrm{high} T}}$ as a function of $d^{*}$.  For $d^{*} \rightarrow 0$, $\gamma(d^{*}) \rightarrow 1$, since in this limit of perfect resolution, each patch appears only once due to thermal fluctuations.  On the other extreme, for $d^{*} \rightarrow \infty$, $\gamma(d^{*}) \rightarrow 1$ as well, since in this limit all patches are congruent.  For other values of $d^{*}$, $\gamma(d^{*}) > 0$ and is expected to have a single maximum - this is where the ``contrast'' is greatest.
As seen in the inset to Fig.~\ref{fig:entropy}, a maximum is found at $d^* \simeq 0.425 \sigma$ (where $\sigma$ is the particle diameter); this value does indeed correspond to a mean displacement at a time just after the plateau~\cite{Sausset:2008d}.
The results we present here use this value of $d^*$ for a density $\rho \sigma^2 \simeq 0.85$. In addition, we use an angular resolution $\delta\alpha = \frac{\sigma}{40} \frac{1}{2 \pi R}$ which appears sufficient to approximate the $\delta\alpha \to 0$ limiting case well.

Fig.~\ref{fig:entropy} shows the patch entropy {\it vs.} patch size for various temperatures.  The entropy exhibits three regimes with increasing $R$: flat (subextensive) at small $R$, followed by a linear rise (the extensive regime), and finally saturation.  The flat regime corresponds to small repetitive patches, the second regime reflects the extensive behavior characteristic of a material disordered on this scale, and the saturation is a finite size effect (Increasing the system size moves the plateau higher).  As discussed above, we extract the patch correlation length $\xi_{S}$ from the correlation volume given by the (inverse) slope of the extensive entropy regime, which, to average out oscillations due to the discrete particle size, we fit with a linear function (see Fig.~\ref{fig:entropy}). We note that the behavior of the entropy for different temperatures clearly indicates that the order is growing as the system is cooled.

In order to determine the relevance of the patch correlation length, it is of interest to compare it to other proposed measures.  To this end, we have computed the point-to-set length $\xi_{\mathrm{PTS}}$~\cite{Bouchaud:2004, Montanari:2006, Biroli:2008} by adapting the method described in~\cite{Biroli:2008} to two-dimensional curved space \footnote {We note that we need to average over $O(100)$ cavities to have good statistics, which is considerably more than in~\cite{Biroli:2008}.}. Details will be published elsewhere, but we find a slightly compressed exponential dependence of the overlap on the size of the cavity that is compatible with the results of~\cite{Biroli:2008}. 
Additionally, we have computed the hexatic ordering length $\xi_{6}$~\cite{Sausset:2010}, which serves as a benchmark for this system.

We present the results in  Fig.~\ref{fig:lengths}, where we plot the various lengths for different temperatures and fragilities.  The three lengths have been linearly rescaled (see the figure caption).  As can be seen, the three lengths show very similar temperature dependence for each of the three fragilities studied, except possibly at the lowest temperatures. The fragility dependence of these length scales (growing faster for larger fragilities) confirms the link between the cooperative character of relaxation and fragility. It is worth noting that the patch entropy method is in this case the least demanding in terms of computer resources, which is why it is computed down to lower temperatures.

\begin{figure}
	\centering
		\includegraphics[width=8 cm,trim= 1cm 0cm 0.5cm 0cm]{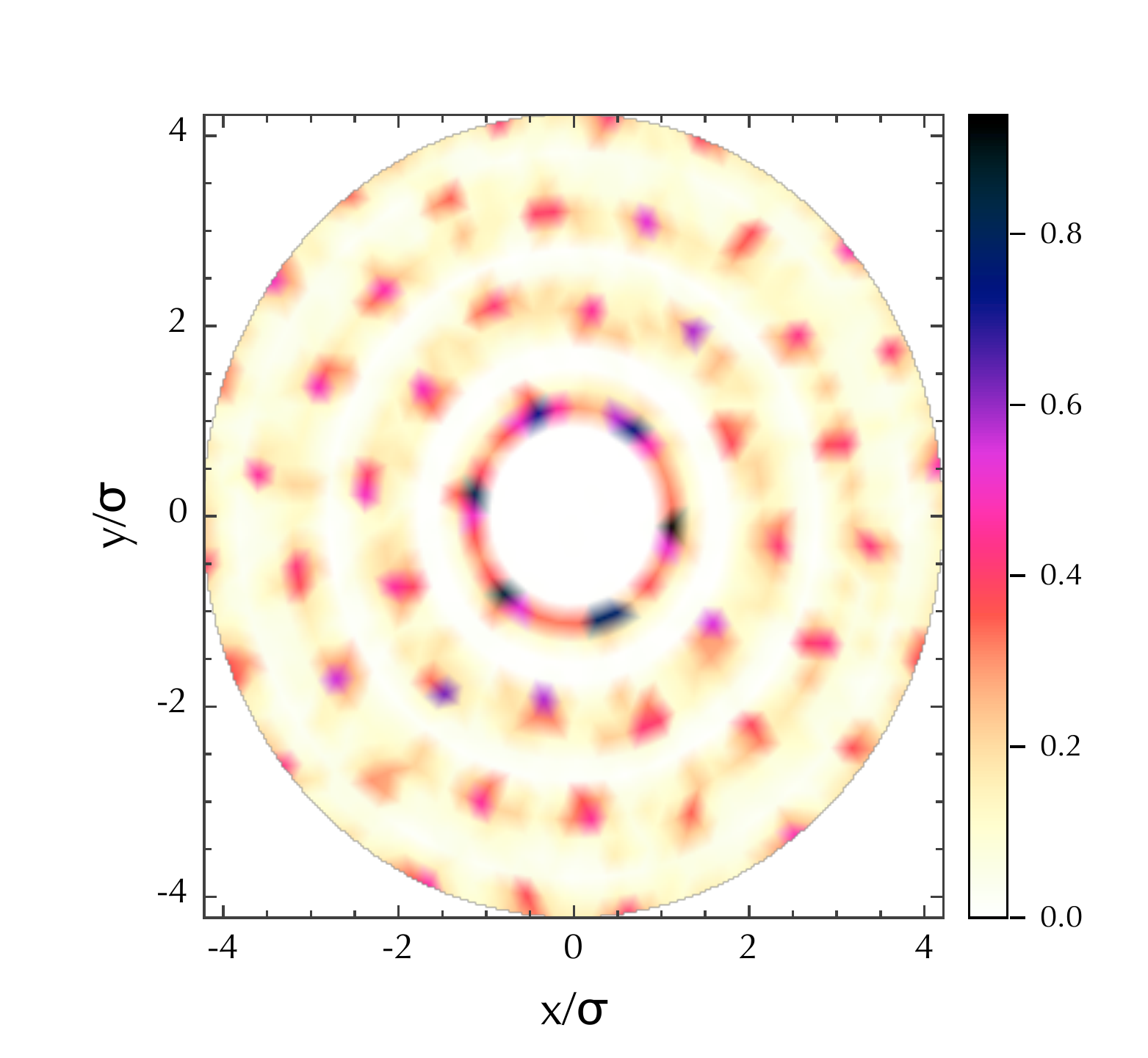}
\caption{Density profile of the mean patch of the largest class of patches: the diameter is equal to $8.4 \sigma$, $T/T^*=0.52$, and $\kappa\sigma = 0.05$. This class contains 1591 repeating patches out of  a total of 4279 patches in the system, so roughly one third of the system is used to construct this density profile.  Red (dark) spots indicate the most probable positions of particles inside of patch of this diameter, indicating a pronounced hexagonal order out to this scale.}
\label{fig:patch}
\end{figure}

An additional advantage of the patch entropy method is that it can be used to extract information about the way in which the system orders.  The idea is that the local order should be reflected in the largest classes of congruent patches, and the associated ``mean patches'' should exhibit the symmetries characteristic of patches in this class. To do this, we construct mean patches from the various equivalence classes computed for a given patch size from a given sample.  This is done by superimposing and adding all the patches in the class (each at its optimal angle).  This creates a density profile which reflects the symmetry of the class.  Clearly this is relevant only if the population of the class is non trivial. 
In Fig.~\ref{fig:patch}, we reconstruct the mean patch of the largest class of patches for a diameter $2R$ smaller than $\xi_S$ in the deeply supercooled regime.  The density profile clearly shows modulations indicating that the most probable particle locations are clearly arranged on a hexagonal lattice. For larger patches, the mean patch begins to blur, indicating that such patches are larger than the structural correlation length. From this, one  could compute an absolute value of the correlation length by looking at the crossover between peaked and blurred mean patches. For instance in Fig.~\ref{fig:patch}, the patch begins to blur for a diameter just above $8.4 \sigma$ (the one of the displayed patch). Thus $\xi_S\simeq 9-10 \, \sigma$ in this case, which seems inconsistent with results from Fig.~\ref{fig:lengths}: at this frustration and temperature $\xi_S\simeq 25 \sigma$. This may be rationalized by noticing that $\xi_S$ is, as stated above, defined up to a multiplicative constant and is linearly rescaled to match $\xi_6$ in Fig.~\ref{fig:lengths}, showing a comparable temperature dependence but implying nothing about the absolute value of the lengths.

Although local hexagonal order was expected for our system, these results  demonstrate the feasibility of the patch entropy method as a way of identifying local order.  It should hold equally well for other amorphous systems where the nature of the order is not known {\it a priori}.  We are currently using this method to elucidate the symmetries of three-dimensional colloidal systems.

\begin{acknowledgments}
	We would like to thank J. Kurchan for fruitful discussions and G. Biroli and G. Tarjus for their critical reading of the manuscript. DL gratefully acknowledges support from Israel Science Foundation grant 1574/08 and US-Israel Binational Science Foundation grant 2008483.
\end{acknowledgments}

\end{document}